\documentclass[reprint,superscriptaddress,showpacs,preprintnumbers,amsmath,amssymb,aps,prl]{revtex4-1}

\usepackage{amssymb}
\usepackage{amsmath}
\usepackage{graphicx}
\usepackage{dcolumn}
\usepackage{bm}
\usepackage{multirow}

\begin{document}

\newcommand{\stat}{5.4\times10^{-7}}
\newcommand{\etav}{(0.3\pm 5.4)\times 10^{-7}}
\newcommand{\ket}[1]{$\left|#1\right\rangle$}
\newcommand{\Rbb}{$^{85}$Rb}
\newcommand{\Rb}{$^{87}$Rb}
\newcommand{\K}{$^{39}$K}
\newcommand{\Eot}{$\eta_{\text{Rb,K}}$}
\newcommand{\EotAB}{$\eta_{\text{A,B}}$}
\newcommand{\EotDelta}{$\Delta\eta$}
\newcommand{\Eotsigma}{$\delta\eta$}
\newcommand{\spaceminus}{$\thinspace -\thinspace$}
\newcommand{\Q}[1]{$Q^{'1}_{\text{#1}}$}
\newcommand{\QQ}[1]{$Q^{'2}_{\text{#1}}$}
\newcommand{\fbminus}[1]{\mbox{$f_{\beta^{e+p-n}_{\text{#1}}}$}}
\newcommand{\fbbarminus}[1]{\mbox{$f_{\beta^{\bar{e}+\bar{p}-\bar{n}}_{\text{#1}}}$}}
\newcommand{\fbplus}[1]{\mbox{$f_{\beta^{e+p+n}_{\text{#1}}}$}}
\newcommand{\fbbarplus}[1]{\mbox{$f_{\beta^{\bar{e}+\bar{p}+\bar{n}}_{\text{#1}}}$}}

\renewcommand{\arraystretch}{1.2}

\title{Quantum Test of the Universality of Free Fall}

\author{D. Schlippert}
\author{J. Hartwig}
\author{H. Albers}
\author{L. L. Richardson}
\author{C. Schubert}
\affiliation{Institut f\"ur Quantenoptik and Centre for Quantum Engineering and Space-Time Research \textnormal{(QUEST)}, Leibniz Universit\"at Hannover, Welfengarten 1, D-30167 Hannover, Germany}
\author{A. Roura}
\affiliation{Institut f\"ur Quantenphysik and Center for Integrated Quantum Science and Technology \textnormal{(IQ$^{\text{ST}}$)}, Universit\"at Ulm, Albert-Einstein-Allee 11, D-89081 Ulm, Germany}
\author{W. P. Schleich}
\affiliation{Institut f\"ur Quantenphysik and Center for Integrated Quantum Science and Technology \textnormal{(IQ$^{\text{ST}}$)}, Universit\"at Ulm, Albert-Einstein-Allee 11, D-89081 Ulm, Germany}
\affiliation{Texas A\&M University Institute for Advanced Study \textnormal{(TIAS)}, Institute for Quantum Science and Engineering \textnormal{(IQSE)}, and Department of Physics and Astronomy, Texas A\&M University, College Station, Texas 77843-4242, USA}
\author{W. Ertmer}
\author{E. M. Rasel}\email{rasel@iqo.uni-hannover.de}
\affiliation{Institut f\"ur Quantenoptik and Centre for Quantum Engineering and Space-Time Research \textnormal{(QUEST)}, Leibniz Universit\"at Hannover, Welfengarten 1, D-30167 Hannover, Germany}

\date{\today}

\begin{abstract}
We simultaneously measure the gravitationally-induced phase shift in two Raman-type matter-wave interferometers operated with laser-cooled ensembles of \Rb~and \K~atoms. Our measurement yields an E\"otv\"os ratio of \Eot~$=\etav$. We briefly estimate possible bias effects and present strategies for future improvements.
\end{abstract}

\pacs{37.25.+k, 03.75.Dg, 04.80.Cc, 06.30.Gv}

\maketitle

\begin{figure}[t]
\begin{center}
\includegraphics[width=.7\linewidth]{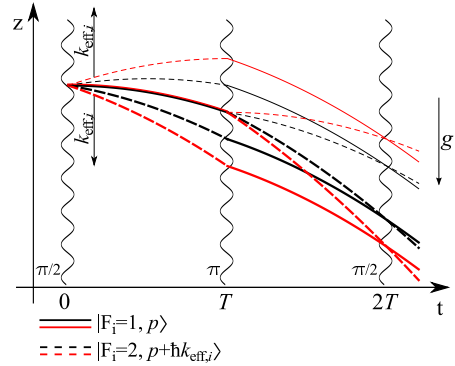}
\caption{(Color online). Space-time evolution of the rubidium and potassium atoms in a Mach-Zehnder-type interferometer positioned in a constant gravitational field pointing downwards. Coherent Raman processes at $t=0,T$ and $2T$ between the states \ket{F_i=1, p} and \ket{F_i=2, p+\hbar k_{\text{eff},i}}, where $i$ is either Rb or K, resulting from the $\pi/2$-, $\pi$- and $\pi/2$-pulses allow for momentum transfer in the downward (thick lines) and the upward (thin lines) direction. The difference in the velocity change between rubidium (black lines) and potassium (red lines) is not to scale.}
\label{mz}
\end{center}
\end{figure}

The universality of free fall (UFF) emerges~\cite{Misner73Freeman} from the equality of the inertial and the gravitational mass, which Heinrich Hertz~\cite{Hertz99} already in 1884 called a "wonderful mystery". In 1915 Albert Einstein made this postulate into one of the cornerstones of general relativity. Although UFF has been verified in numerous tests~\cite{Williams04PRL,Schlamminger08PRL} today different scenarios reconciling general relativity and quantum mechanics allow a violation of the UFF. For this reason more precise tests are presently pursued~\cite{Touboul12CQG,Freire12CQG,Ransom14Nature} and new measurement techniques are developed. One intriguing approach consists of comparing the accelerations of different quantum objects to a high precision. In this Letter, we report the first quantum test of the UFF with matter waves of two different atomic species.\\\indent
We simultaneously compare the free-fall accelerations $g_\text{Rb}$ and $g_\text{K}$ of \Rb~and \K~measured by inertial-sensitive Mach-Zehnder type interferometers shown in \mbox{Fig. \ref{mz}} employing stimulated two-photon Raman transitions and extract the E\"otv\"os ratio
\begin{equation}\label{eotvos}
\eta_{\text{Rb,K}}\equiv 2\thickspace\frac{g_{\text{Rb}}-g_{\text{K}}}{g_{\text{Rb}}+g_{\text{K}}}
=2\thickspace\frac{\left(\frac{m_{\text{gr}}}{m_{\text{in}}}\right)_{\text{Rb}}
-\left(\frac{m_{\text{gr}}}{m_{\text{in}}}\right)_{\text{K}}}
{\left(\frac{m_{\text{gr}}}{m_{\text{in}}}\right)_{\text{Rb}}
+\left(\frac{m_{\text{gr}}}{m_{\text{in}}}\right)_{\text{K}}}.
\end{equation}
A UFF violation, that is, \Eot~$\neq 0$ yields a difference in the inertial mass $m_{in}$ and gravitational mass $m_{gr}$ of, or an additional force coupling differently to the two species.\\\indent
There exist two types of quantum tests of the UFF: (i) The first one~\cite{Peters99Nature,Merlet10Metrologia} compares the accelerations obtained by measuring the gravitationally-induced phase shift of freely falling matter waves of neutrons~\cite{Colella75PRL}, or atoms to the one measured with classical gravimeters. (ii) The second one which is solely of quantum nature compares this phase shift for two types of matter waves such as different rubidium isotopes~\cite{Fray04PRL,Fray09SSRev,Bonnin13PRA} or strontium isotopes~\cite{Tarallo14arxiv}. Today, there are numerous initiatives~\cite{Varoquaux09NJP,Dickerson13PRL,Sugarbaker13PRL,VanZoest10Science,Muentinga13PRL,
Mueller13arxiv,Aguilera13arxiv,Schubert13arxiv} on the way to test the UFF with matter-wave interferometers both on ground and in microgravity.\\\indent
\begin{table*}
\caption{Comparison of test masses A and B employed in different tests of the UFF with respect to their effective charges \Q{X}, \QQ{X}~and \fbplus{X}, \fbminus{X}, \fbbarminus{X}, \fbbarplus{X}~where X is either A or B calculated according to \cite{Damour12CQG} and \cite{Hohensee13PRL}, respectively, using nuclide data from \cite{Audi03}. A larger absolute number corresponds to a larger anomalous acceleration and thus higher sensitivity to violations of the EEP. For Ti we assume a natural occurrence of isotopes~\cite{Laeter09}.}
\begin{ruledtabular}
\begin{tabular}{ c c c | d d  d d d d }
  \multirow{2}{*}{A}& \multirow{2}{*}{B}&\multirow{2}{*}{Ref.} &\multicolumn{1}{c}{\Q{A}\spaceminus\Q{B}}&\multicolumn{1}{c}{\QQ{A}\spaceminus\QQ{B}}& \multicolumn{1}{c}{\fbminus{A}\spaceminus\fbminus{B}} & \multicolumn{1}{c}{\fbplus{A}\spaceminus\fbplus{B}} & \multicolumn{1}{c}{\fbbarminus{A}\spaceminus\fbbarminus{B}} & \multicolumn{1}{c}{\fbbarplus{A}\spaceminus\fbbarplus{B}} \\
  &&&\multicolumn{1}{c}{$\cdot 10^4$}&\multicolumn{1}{c}{$\cdot 10^4$}&\multicolumn{1}{c}{$\cdot 10^2$}&\multicolumn{1}{c}{$\cdot 10^4$}&\multicolumn{1}{c}{$\cdot 10^5$}&\multicolumn{1}{c}{$\cdot 10^4$}\\
\hline

\textsuperscript{9}Be& Ti&\cite{Schlamminger08PRL}  &-15.46 &-71.20& 1.48 &-4.16 & -0.24 &-16.24\\

\Rbb&\Rb&\cite{Fray04PRL,Fray09SSRev,Bonnin13PRA} &0.84& -0.79 &-1.01& 1.81 &1.04& 1.67\\

\textsuperscript{87}Sr&\textsuperscript{88}Sr &\cite{Tarallo14arxiv} &0.42& -0.39 &-0.49& 2.04 &10.81& 1.85\\

\K&\Rb&[This work]& -6.69& -23.69& -6.31& 1.90& -62.30& 0.64\\
\end{tabular}
\label{violation}
\end{ruledtabular}
\end{table*}
Matter-wave tests of the UFF differ from their classical counterparts in several aspects: (i) The coherence lengths of these quantum objects differ~\cite{Goeklue08CQG} by orders of magnitude as compared to classical ones. (ii) Matter waves allow us to perform both, tests of the redshift and of the free fall using the same species. (iii) Quantum tests are performed with spin-polarized ensembles, a feature that is only available in few specific scenarios outside of matter-wave tests~\cite{Tarallo14arxiv,Leitner64PRL,Laemmerzahl98proceedings,Hsieh89ModPL}. (iv) Experiments with matter waves take advantage of chemical species of highest isotopic purity, and (v) quantum tests enlarge the set of test mass pairs employed for example in torsion-balance experiments, which are mostly performed with non-magnetic, conducting solids~\cite{Adelberger09}.\\\indent
Indeed, depending on the models for violations of UFF and the bounds on them derived from tests, the different combinations of test materials act as a sensitivity lever, and provide complementary information. In this way new combinations of test masses impose different constraints on the model~\cite{Damour12CQG,Hohensee13PRL}. \mbox{Table \ref{violation}} compares four choices of test masses employed (i) in the best torsion balance experiment, \textsuperscript{9}Be~vs. Ti~\cite{Schlamminger08PRL}, (ii) in the quantum tests, \Rb~vs. \Rbb~\cite{Fray04PRL,Fray09SSRev,Bonnin13PRA} or \textsuperscript{87}Sr vs. \textsuperscript{88}Sr~\cite{Tarallo14arxiv}, (iii) and in our experiment, \Rb~vs. \K, with respect to their sensitivities to possible violations of the Einstein equivalence principle (EEP) predicted by the dilaton model~\cite{Damour12CQG}, and the standard-model extension~\cite{Hohensee13PRL}.\\\indent
In the dilaton model different forces may act on neutrons and protons and we can attribute effective charges \Q{X}~and\QQ{X}~to the individual species X according to its composition. They can be calculated, and by determining the E\"otv\"os ratio~\cite{Damour12CQG}
\begin{equation}\label{damour}
\eta_{\text{A,B}}~\widetilde =~ D_1(Q^{'1}_{\text{A}}-Q^{'1}_{\text{B}})+D_2(Q^{'2}_{\text{A}}-Q^{'2}_{\text{B}})
\end{equation}
they set bounds on the violation parameters $D_1$ and $D_2$. Here, a larger difference in the effective charge corresponds to a larger anomalous acceleration in an EEP-violating scenario. Vice versa, a test ruling out a violation of the EEP at a certain level imposes tighter bounds on the violation coefficients.\\\indent
An expression similar to \mbox{Eq. (\ref{damour})} exists for the standard-model extension. Here, the E\"otv\"os ratio can be written as
\begin{equation}
\eta_{\text{A,B}}~\widetilde = ~\beta_{\text{A}}-\beta_{\text{B}}
\end{equation}
with violation parameters
\begin{equation}
\begin{aligned}
\beta_\text{X}\equiv \thickspace f_{\beta^{e+p-n}_{\text{X}}}\beta^{e+p-n}+f_{\beta^{e+p+n}_{\text{X}}}\beta^{e+p+n}\\
+f_{\beta^{\bar{e}+\bar{p}-\bar{n}}_{\text{X}}}\beta^{\bar{e}+\bar{p}-\bar{n}}
+f_{\beta^{\bar{e}+\bar{p}+\bar{n}}_{\text{X}}}\beta^{\bar{e}+\bar{p}+\bar{n}}
\end{aligned}
\end{equation}
for species X~\cite{Hohensee13PRL}.\\\indent
\begin{figure*}[t]
\begin{center}
\includegraphics[width=.9\linewidth]{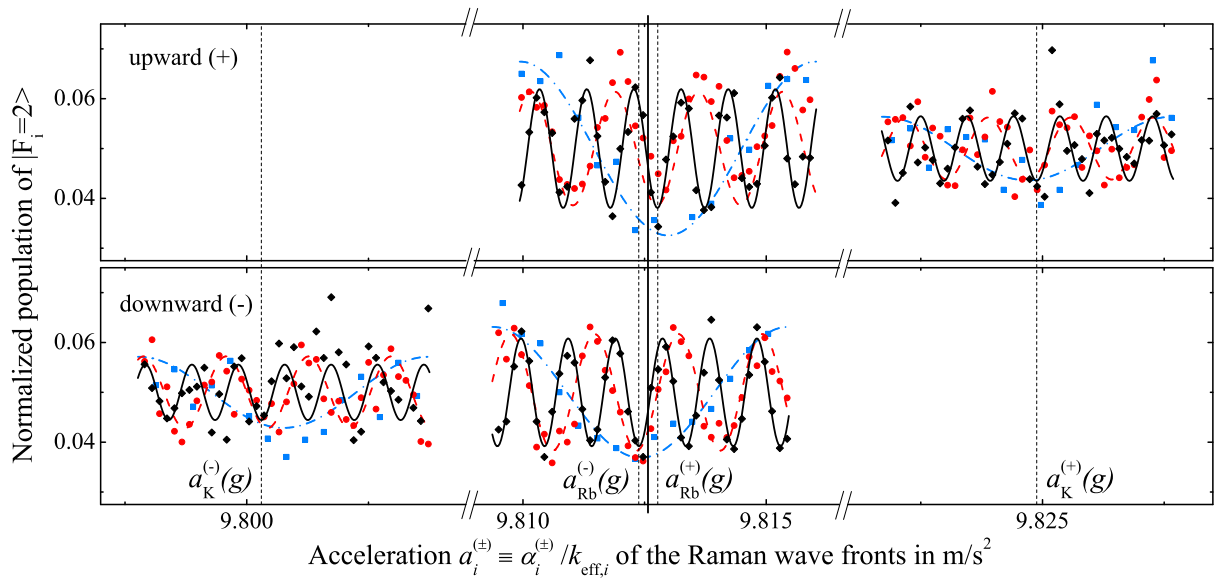}
\caption{(Color online).
Determination of the gravitational accelerations $g_{\text{Rb}}$ and $g_{\text{K}}$ of rubidium and potassium from two orientations of the interferometer determined by the upward $(+)$ and downward $(-)$ direction of the momentum transfer indicated in Fig. \ref{mz}. We display the corresponding signals (individual data points) and the respective sinusoidal least squares fit functions (curves) for three pulse separation times $T=8~$ms (blue squares, dash-dotted blue line), $T=15~$ms (red circles, dashed red line), and $T=20~$ms (black diamonds, solid black line). Due to bias contributions that are independent of the direction of the momentum transfer the values $a^{(\pm)}_i(g)$ are offset symmetrically around their half difference $(a^{(+)}_i(g)-a^{(-)}_i(g))/2$ marked for rubidium (solid vertical line). The displayed datasets have been adjusted in their signal offsets. Moreover, we have subtracted a linear fitting function from the datasets to correct for slow offset drifts in the detection caused by variations in the arrival time.}
\label{fringes}
\end{center}
\end{figure*}
Tests based on different rubidium or strontium isotopes may make up for their similar response with respect to both violation scenarios apparent from \mbox{Table \ref{violation}} by their strong common mode rejection of systematic errors and correlated noise sources~\cite{Bonnin13PRA}. Similar to \textsuperscript{9}Be~vs. Ti, our test-mass pair displays a higher sensitivity with respect to the dilaton model~\cite{Damour12CQG}. In addition, the strontium isotopes and, in particular, our choice of \Rb~and \K~are complementary to existing torsion balance tests with respect to the parameters \fbminus{A}\spaceminus\fbminus{B} (\fbbarminus{A}\spaceminus\fbbarminus{B}) and \fbplus{A}\spaceminus\fbplus{B} (\fbbarplus{A}\spaceminus\fbbarplus{B}) reflecting the (anti)neutron excess and the total (anti)baryon number charges~\cite{Hohensee13PRL} of the particles, respectively.\\\indent
Our magnetically-shielded apparatus features a dual-species magneto-optical trap (MOT) which is loaded from a cold atomic beam formed by transverse laser cooling. Since the D$_2$ lines of \Rb~(780~nm) and \K~(767~nm) have a relatively small difference in wavelength most optics can be used for both species. In a typical experimental cycle we load $8\times 10^8$~atoms ($3\times 10^7$~atoms) of \Rb~(\K) into the 3D-MOT within $1~$s. Due to the lighter mass and a small hyperfine splitting of the excited state of only a few linewidths, \K~requires~\cite{Landini11PRA} more complex cooling techniques as compared to \Rb. We thus optimize the experiment in favor of larger \K~atom numbers to better match the performance of both interferometers. After cooling~\cite{Dalibard89JOSAB,Landini11PRA} to temperatures of $27~\mu$K ($32~\mu$K) for \Rb~(\K), we optically pump all atoms into their respective \ket{F=1}-state and release both ensembles into free fall.\\\indent
The free-falling matter waves are coherently split, redirected and recombined by employing stimulated two-photon Raman transitions~\cite{Kasevich91PRL,Peters99Nature} with momentum transfer $k_{\text{eff},i}$ (throughout this Letter $i$ is either Rb or K). The Raman laser light for both species is superimposed on a dichroic mirror with parallel polarizations and guided to the atoms with a polarization-maintaining single-mode fiber. After the fiber the Gaussian-shaped beams are expanded to a waist of $1.4~$cm and circularly polarized by a quarter-wave plate before traversing the vacuum chamber along the vertical axis. Below the chamber the beam is retroreflected by a mirror (specified peak-to-valley flatness: $\lambda/20$) mounted on a commercial vibration isolation platform. Thus, two pairs of counter-propagating beams are generated.\\\indent
Due to the narrow transition resonance the two pairs of Raman beams allow us to scatter the atoms either upwards or downwards, and hence to invert the orientation of the interferometer with respect to the direction of the free fall \mbox{(Fig. \ref{mz})}.\\\indent
Both, the potassium and the rubidium interferometer, are operated simultaneously at laser detunings of $\Delta_{\text{Rb}}=-1.6~$GHz ($\Delta_{\text{K}}=-3.3~$GHz) for rubidium (potassium) with identical pulse widths $\tau_{\pi}=15~\mu$s and pulse separation times $T=20~$ms. Normalized signals are obtained from the two output ports via state-selective fluorescence detection. A full experimental cycle takes about $1.6~$s.
In order to extract the value of the acceleration $g_i$ from the leading order phase shift
\begin{equation}\label{phaseshift}
\phi_i=(k_{\text{eff},i}\cdot g_i-2\pi\cdot\alpha_i)\cdot T^2
\end{equation}
the Raman lasers are chirped~\footnote{Chirping the laser frequency mimics an acceleration of the beam splitting light inertial frame. It should be noted that this null measurement relies on the validity of Einstein's principle of equivalence for light waves, stating that locally we cannot distinguish gravity from an acceleration.} at a rate $\alpha_i$ which translates into an acceleration $a_i\equiv \alpha_i/k_{\text{eff},i}$ of the laser fields. The resulting count rates at the two exit ports of the interferometer oscillate as a function of $\alpha_i$.\\\indent
According to Eq. (\ref{phaseshift}), the phase shift $\phi_i$ scales quadratically with the pulse separation time $T$ and allows us in this way obtain $g_i$. Indeed, in the absence of perturbations the count rate is independent of $T$, when the wave acceleration $a_i$ determined the chirp rate $\alpha_i$ compensates exactly the free-fall rate, that is, $\alpha_i(g_i)/k_{\text{eff},i}\equiv g_i$.\\\indent
However, perturbations shift the fringes and require a more complex measurement scheme. For this reason we obtain fringe patterns for both, the upward- and the downward-oriented interferometers determined by the direction of momentum exchange shown in Fig. \ref{mz}. In contrast to the gravitational phase shift, the most important non-inertial phase shifts in our dual species interferometers do not change their sign when inverting the interferometer and thus are canceled in the half-difference $(a_i^{(+)}(g)-a_i^{(-)}(g))/2$ of the signals~\cite{McGuirk02PRA,Louchet-Chauvet11NJP}.\\\indent
In turn, phase shifts due to perturbations which switch their sign in the inverted interferometer originating for example from the two-photon light shift or curved wave fronts of the beam-splitting light field have to be analyzed by other means for the uncertainty budget and are listed in \mbox{Table~\ref{systematics}}.\\\indent
In order to perform a test of the UFF data was taken over $\sim 4$~h by continuously tuning the ramp rates of rubidium and potassium around their central fringe positions $a^{(\pm)}_i(g)$ with alternating orientations of the interferometer. As shown in \mbox{Fig. \ref{fringes}} the contrast of the resulting interference patterns is presently at the few percent level and can be further enhanced by additional cooling and state selection and by employing a dipole trap common to both species~\cite{Zaiser11PRA}. The E\"otv\"os ratio \Eot~can then be calculated from the obtained single-species signals using \mbox{Eq. (\ref{eotvos})}.\\\indent
The Allan deviations~\cite{Allan66} shown in \mbox{Fig. \ref{adev}} correspond to the statistical uncertainty of the normalized single-species acceleration-signals $g_i$ and the resulting E\"otv\"os ratio defined by \mbox{Eq. (\ref{eotvos})}. We attribute the relatively low short-term stability of potassium to the technical noise which currently limits our measurement. After $4096$~s of integration we achieve a statistical uncertainty of $\sigma_\eta=\stat$ in our determination of the E\"otv\"os ratio.\\\indent
\begin{figure}
\begin{center}
\includegraphics[width=.9\linewidth]{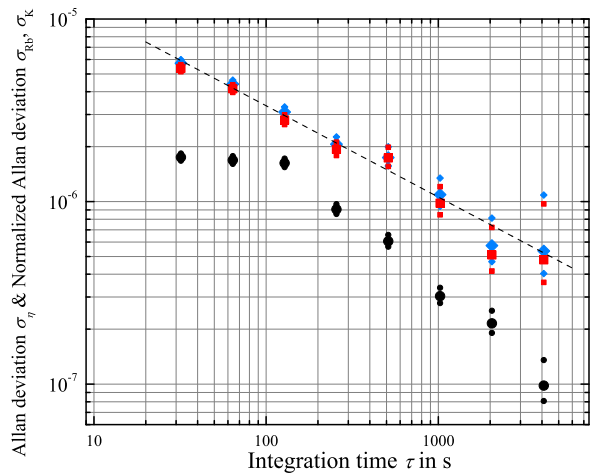}
\caption{(Color online). Normalized Allan deviations $\sigma_{\text{Rb}}$, $\sigma_{\text{K}}$, and $\sigma_\eta$ of the signals providing us with the accelerations $g_{\text{Rb}}$ and $g_{\text{K}}$ of rubidium (black circles) potassium (red squares), and of the E\"otv\"os ratio \Eot~(blue diamonds) and its asymptotic behavior (dashed line), respectively, in their dependence on the integration time $\tau$. We achieve a statistical uncertainty of the E\"otv\"os ratio of $\sigma_\eta=\stat$ after $4096~$s integration. The measurement is solely limited by the stability of the potassium signal. The first data point at $\tau=32~$s is given by the time required to obtain a single data point with the scheme described in detail in the text.}
\label{adev}
\end{center}
\end{figure}
\mbox{Table~\ref{systematics}} summarizes estimates for the remaining bias contributions \EotDelta~in our measurement of the E\"otv\"os ratio using the rubidium-potassium combination and their uncertainties \Eotsigma. Our assessment is based on the experimental parameters described above and a measured residual magnetic field gradient. Our theoretical model identifies the quadratic Zeeman effect and the wave front curvature of the Raman beams as the leading effects for systematic biases. Taking into account all bias contributions we can infer an E\"otv\"os ratio \Eot~$=\etav$.\\\indent
\begin{table}
\begin{ruledtabular}
\caption{Estimates for several bias contributions \EotDelta~to the E\"otv\"os ratio and their uncertainties \Eotsigma~(for the use of an optical dipole trap we expect the improved uncertainties $\delta\eta^{\text{adv}}$). We treat the uncertainties to be uncorrelated at the level of accuracy.}
\begin{tabular}{ p{3.5cm} | r@{$\times$}l  r@{$\times$}l | r@{$\times$}l}

Contribution &\multicolumn{2}{c}{\EotDelta} & \multicolumn{2}{c}{\Eotsigma} &\multicolumn{2}{c}{$\delta\eta^{\text{adv}}$}\\
\hline

2\textsuperscript{nd} order Zeeman effect &-5.8&$10^{-8}$&2.6&$10^{-8}$& 3.0&$10^{-9}$\\

Wave front abberation  & \multicolumn{2}{c}{0}&1.2&$10^{-8}$& 3.0&$10^{-9}$\\

Coriolis force & \multicolumn{2}{c}{0}&9.1&$10^{-9}$& 1.0&$10^{-11}$\\

Two-photon light shift &3.7&$10^{-9}$&7.3&$10^{-11}$& 7.3&$10^{-11}$\\

Effective wave vector & \multicolumn{2}{c}{0}&1.3&$10^{-9}$& 1.3&$10^{-9}$\\

Gravity gradient (1\textsuperscript{st} o.)& 9.5&$10^{-11}$&9.5&$10^{-12}$& 1.0&$10^{-13}$\\
\hline
Total &-5.4&$10^{-8}$ &3.1&$10^{-8}$ &4.4&$10^{-9}$\\

\end{tabular}
\label{systematics}
\end{ruledtabular}
\end{table}
In conclusion, we have demonstrated the first quantum test of the UFF using matter waves of two different neutral elements. State preparation in a dipole rather than a magneto-optical trap will alleviate systematic effects by allowing for lower expansion rates and better magnetic field characterization. Based on these improvements, we project~\footnote{The extrapolation for $\delta\eta^{\text{adv}}$ is based on the following parameters for the atomic ensembles and the trap: A temperature of the Rb-K mixture of $10~\mu$K, a horizontal trap alignment uncertainty of $1.5^{\circ}$, $3$~ms trap switch-off time and an uncertainty of the trap power before release of $0.2$\%.} with our apparatus ppb-level tests within a few thousand seconds integration time as indicated by the last column of Table \ref{systematics}.\\\indent
Our experiments open up a wide range of new experimental and theoretical studies. Apart from such tests being performed in future with large scale fountains~\cite{Dickerson13PRL,Sugarbaker13PRL} and in microgravity environments~\cite{VanZoest10Science,Muentinga13PRL,Aguilera13arxiv,Schubert13arxiv}, we anticipate a large interest in a joint analysis of measurements performed with different species and isotopes in the elaborate theoretical scenarios~\cite{Aguilera13arxiv,Damour12CQG}. For example, in the frame of the standard-model extension a Rb-K comparison at a level of $10^{-11}$ will improve the global bounds on the EEP violation parameters for neutral matter by two orders of magnitude~\cite{Mueller13arxiv}. Moreover, depending on the chosen methods, quantum objects allows to study different aspects emerging from quantum mechanics rather than from classical systems~\cite{Jenke11NaturePhys,Kajari10APB}. Recent proposals for tests with bosonic and fermionic matter as well as with matter in superposition states~\cite{Zych11NatureComms,Zych14DPG} represent only two of the many examples demonstrating the potential for future extensions of our experiment.
\begin{acknowledgments}
We thank \mbox{M. A. Hohensee} and \mbox{H. M\"uller} for their theory support with respect to the standard-model extension, and \mbox{F. Pereira dos Santos}, \mbox{H. Ahlers},  \mbox{N. Gaaloul}, \mbox{W. Herr}, \mbox{P. Hamilton}, and \mbox{C. Klempt} for fruitful discussions. \mbox{M. Zaiser}, \mbox{D. Tiarks}, \mbox{U. Velte}, and \mbox{C. Meiners} contributed valuably during the early stages of the experiment. This work was financed by the German Research Foundation (DFG) via the Cluster of Excellence Centre for Quantum Engineering and Space-Time Research (QUEST), and the German Space Agency (DLR) with funds provided by the Federal Ministry of Economics and Technology (BMWi) due to an enactment of the German Bundestag under Grant No. DLR 50WM1142 (project PRIMUS-II). WPS gratefully acknowledges the support by a Texas A\&M University Institute of Advanced Study Faculty Fellowship.
\end{acknowledgments}

\bibliography{QTest_UFF_resubmission.bbl}

\end{document}